# Abruptly Focusing and Defocusing Needles of Light and Closed-Form Electromagnetic Wavepackets

Liang Jie Wong[1,*] and Ido Kaminer[2]

[1]Singapore Institute of Manufacturing Technology, 2 Fusionopolis Way, Innovis, Singapore 138634

[2]Department of Physics, Massachusetts Institute of Technology, 77 Massachusetts Avenue, Cambridge 02139, Massachusetts, USA

*wonglj@simtech.a-star.edu.sg

**Abstract:** Fourier optics enforces a tradeoff between length and narrowness in electromagnetic wavepackets, so that a narrow spatial focus diffracts at a large divergence angle, and only infinitely wide beams can remain non-diffracting. We show that it is possible to bypass this tradeoff between the length and narrowness of intensity hotspots, and find a family of electromagnetic wavepackets that abruptly focus to and defocus from high-intensity regions of any aspect ratio. Such features are potentially useful in scenarios where one would like to avoid damaging the surrounding environment, for instance, to target tumors very precisely in cancer treatment, drill holes of very precise dimensions in laser machining, or trigger nonlinear processes in a well-defined region. In the process, we also construct the first closed-form solutions to Maxwell's equations for finite-energy electromagnetic pulses. These pulses also exhibit intriguing physics, with an on-axis intensity peak that always travels at the speed of light despite inherent diffraction.

The ability to tailor the shape of electromagnetic fields plays an important role in contemporary physics research due to its potential to transform the scientific and technological landscape. Accelerating beams [1, 2], abruptly autofocusing beams [3], beams with topological charge [4-6], needle beams [7], and beams containing intricate vortex and field-line loops [8-9] exhibit fascinating physics and have applications that range from materials processing to medical care [10-18]. Long, narrow, and well-localized hotspots are potentially very useful for applications like high-aspect-ratio laser drilling of features such as microfluidic channels. In many types of focused electromagnetic wavepackets, however, a fundamental constraint prevents one from realizing high intensity regions that are arbitrarily long and narrow. This constraint arises from Fourier optics (equivalently, the Heisenberg Uncertainty Principle if the wavepacket describes a single photon wavefunction), which relates the spatial spread of a wavepacket in a transverse dimension (say $x$) to its angular spread in the corresponding wavevector ($k_x$) [19, 20]

$$\Delta x \Delta k_x \geq 1/2, \qquad (1)$$

where $\Delta$ denotes the standard deviation of the associated variable. Eq. (1) correlates a high transverse localization with a large momentum spread that destroys this localization within a short distance of the focus in typical wavepackets. This is the very reason a single-slit diffraction experiment, for instance, produces an interference pattern that is much broader when the slit is narrower [19, 20]. As Fig. 1a and b illustrate, a tradeoff thus exists between the narrowness and length of the high intensity region in typical electromagnetic wavepackets.

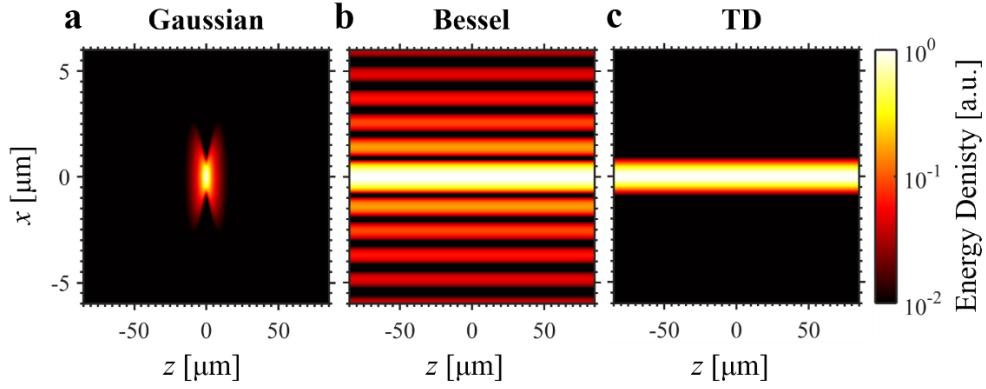

Figure 1. In Fourier optics, a fundamental tradeoff exists between transverse confinement and defocusing length for monochromatic wavepackets as well as many types of electromagnetic pulses. To exemplify this, the locally-time-averaged energy densities of a linearly-polarized Gaussian beam and Bessel beam are shown in (a) and (b) respectively. A small spot size (1.22 μm diameter, wavelength $\lambda_0 = 0.8$ μm) in a Gaussian beam (a) entails a small confocal parameter, whereas the infinitely long Bessel profile (b) entails an infinite width inasmuch as it contains infinite power in its transverse lobes. In temporally diffracting (TD) electromagnetic wavepackets, whose locally-time-averaged energy density is shown in (c) for the case of linear polarization, the tradeoff between transverse confinement and defocusing length does not exist. Instead, waves of different frequencies, are superposed so that a different Fourier transform limit leads to a tradeoff between transverse confinement and pulse duration, making any finite length:width aspect ratios (here, 50,000) possible. This method allows direct control over the spatial profile of the focus without being constrained by Fourier optics. The aspect ratio at any $\Delta\omega/\omega_0$ can be controlled by adjusting the spread in $k_z$ (a larger aspect ratio entails a smaller $k_z$ spread). The energy of the pulse remains finite for any finite spread in $k_z$. Here, peak wavelength $\lambda_0 = 0.8$ μm, i.e. peak frequency $\omega_0 = 2\pi c/\lambda_0 = 2.36 \times 10^{15}$ rad/s and standard deviation $\Delta\omega/\omega_0 = 3.8\%$, implying a pulse duration of 9.7 fs.

Here, we find a way around this fundamental tradeoff by formulating electromagnetic pulses such that the Fourier transform limit correlates transverse localization with time $t$ instead of the propagation spatial dimension $z$. This approach gives us temporally diffracting (TD) electromagnetic wavepackets (Fig. 1c), a special family of pulses that exhibit rich spatiotemporal dynamics and intriguing physics. In particular, they can create high-intensity regions of any finite length:width aspect ratios, with strong transverse confinement (quasi-Gaussian). These pulses feature an on-axis intensity peak that always travels at the speed of light despite inherent diffraction, as well as regimes of abrupt focusing and defocusing that recommends them for applications where very specific regions have to be targeted by intense fields.

Ongoing research into beam and pulse shaping continues to reveal a wealth of specially shaped electromagnetic pulses. These can always be described as a coherent superposition of monochromatic beams of a range of frequencies, with notable families of beams that can be naturally represented as superpositions of Bessel beams [21-22] and remain propagation invariant in space. Other resulting superpositions can create families of X-wave pulses [23,24], needle-like beams and pulses [7,25-30], propagation-invariant pulses [25,26], as well as accelerating [1,2,31-33] and non-accelerating [26] Airy-shaped pulses. In comparison, our wavepackets are the first closed-form description of finite-energy pulses in free space ("closed-form" meaning an explicit analytic expression that require no numerical integrals). Such exact solutions of Maxwell's equations provide new insight into extreme pulse dynamics in highly nonparaxial and short-duration pulses, revealing strong intensity confinements in high aspect ratio hotspots.

To formulate a TD wavepacket, we first consider $\psi = \psi_0(x, y, t, k_z)\exp(ik_z z)$. According to the scalar electromagnetic wave equation $(\nabla^2 - 1/c^2\, \partial_t^2)\psi = 0$ (from $\psi$, vector electromagnetic fields are readily obtained via the Hertz potentials, as discussed later), this wavepacket evolves as

$$\frac{\partial^2 \psi_0}{\partial t^2} = c^2 \nabla_\perp^2 \psi_0 - c^2 k_z^2 \psi_0, \qquad (2)$$

where $\nabla_\perp^2 \equiv \partial_x^2 + \partial_y^2$ and c is the speed of light. Eq. (2) is similar to the Helmholtz equation $\nabla^2 \psi_0 + \omega^2/c^2\, \psi_0 = 0$, except that whereas the Helmholtz equation applies to a monochromatic wavepacket of single angular frequency $\omega$, Eq. (2) applies to a wavepacket containing a single $k_z$ component. A general solution to the electromagnetic wave equation then takes the form

$$\psi(x,y,z,t) = \int_{-\infty}^{\infty} \psi_0(x,y,t;k_z) F(k_z) e^{ik_z z} dk_z, \tag{3}$$

where $F(k_z)$ is an arbitrary distribution in $k_z$. Eqs. (2) and (3) are a re-write of the usual wavepacket formulation method, in which monochromatic components of different frequencies $\omega$ are integrated or summed. However, this simple re-write allows a shift in perspective that simplifies the design of ultrashort and few-cycle pulses as described below.

The regular Helmholtz equation correlates the dimensions of the wavepacket in $x$ and $z$ via the Fourier transform limit, leading to the tradeoff in length and narrowness that makes long and narrow intensity hotspots with strong transverse confinement fundamentally impossible. To bypass this restriction, Eq. (2) correlates the dimensions of the wavepacket in $x$ and $t$ instead, necessitating a short pulse duration – and hence a wide frequency range – to obtain a narrow beam waist, e.g., the wavepacket of 1.2 µm spot size in Fig. 1c has a frequency spread of 3.8%, giving a pulse duration of 9.7 fs. This approach makes it possible to achieve intensity hotspots of very large longitudinal extent and narrow transverse confinement.

We introduce a new closed-form wavepacket solution of Maxwell's equations that illustrate the above properties. It is given by the following scalar expression, from which the vectorial electric and magnetic fields are readily obtained in closed-form via the Hertz potentials, at cylindrical coordinates ($r$, $z$) and time $t$ as

$$\psi = -i \frac{ict + a}{k_0 \tilde{R}^2} \left( \frac{1}{k_0 \tilde{R}} f^{-s-1} + \frac{s+1}{s} f^{-s-2} \right), \tag{4}$$

where $f \equiv 1 - k_0(iz + a - \tilde{R})/s$, $\tilde{R} \equiv [r^2 + (ict + a)^2]^{1/2}$ is the complex length, $\omega_0 = k_0 c = 2\pi c/\lambda_0$ is the central angular frequency of the pulse, and c is the speed of light in the linear, homogeneous, time-invariant and isotropic medium (e.g., free space). To gain some physical intuition, we note that parameters *a* and *s* control the focal spot size and pulse length of the electromagnetic pulse. [34]. Vector solutions of electromagnetic fields **E** and **H** are readily obtained by treating the scalar solution (4) as a component of Hertz vectors $\mathbf{\Pi}_e$ and $\mathbf{\Pi}_m$, and applying the equations [35]

$$\mathbf{E} = \nabla \times \nabla \times \mathbf{\Pi}_e - \mu \frac{\partial}{\partial t} \nabla \times \mathbf{\Pi}_m$$
$$\mathbf{H} = \nabla \times \nabla \times \mathbf{\Pi}_m + \varepsilon \frac{\partial}{\partial t} \nabla \times \mathbf{\Pi}_e \quad , \quad (5)$$

where $\varepsilon$ and $\mu$ are the medium's permittivity and permeability respectively. For instance, the radially-polarized TM10 (Fig. S1.1) and azimuthally-polarized TE10 modes are obtained by setting $\mathbf{\Pi}_e = \psi \hat{z}$, $\mathbf{\Pi}_m = 0$, and $\mathbf{\Pi}_e = 0$, $\mathbf{\Pi}_m = \psi \hat{z}$ respectively. Linearly-polarized fundamental modes (Fig. S1.2) are obtained by setting $\mathbf{\Pi}_e = \psi \hat{x}$, $\mathbf{\Pi}_m = 0$, or $\mathbf{\Pi}_e = 0$, $\mathbf{\Pi}_m = \psi \hat{y}$, or some linear combination thereof. The real electromagnetic fields are then given by Re{**E**} and Re{**H**}. From (5), the energy density of the field at time *t* is given by $u(t) = \left[ \varepsilon \left| \text{Re}\{\mathbf{E}(t)\} \right|^2 + \mu \left| \text{Re}\{\mathbf{H}(t)\} \right|^2 \right] / 2$. For the convenience of visualization, however, we use the locally-time-averaged energy density defined at time *t* as

$$\bar{u}(t) = \frac{1}{4} \left[ \varepsilon \mathbf{E}(t) \cdot \mathbf{E}^*(t) + \mu \mathbf{H}(t) \cdot \mathbf{H}^*(t) \right], \quad (6)$$

which is simply the energy density after averaging out local oscillations due to the oscillatory nature of the electromagnetic field. (The oscillations in some of the cases we study are too numerous to resolve graphically; hence we have adopted (6) instead of the non-averaged version).

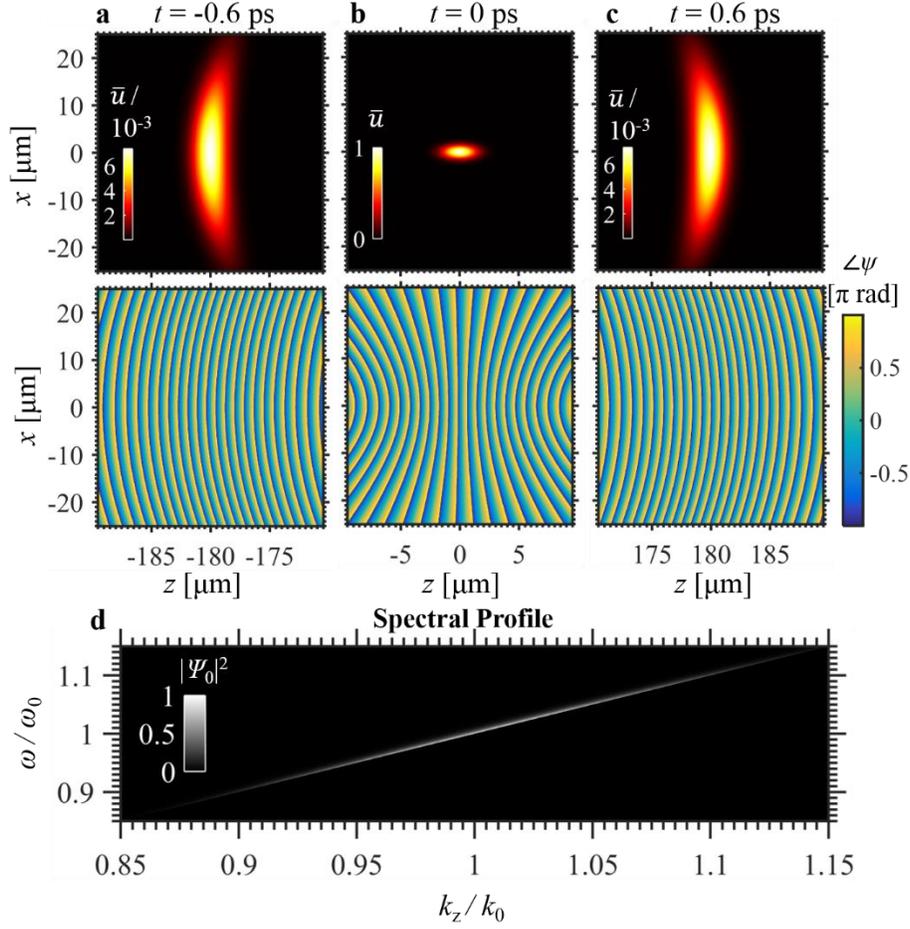

Figure 2. Closed-form, exact description of a tightly-focused, few-cycle linearly-polarized pulse. Snapshots of the propagating pulse are provided at $t = -0.6$ ps (a), $t = 0$ (b) and $t = 0.6$ ps (c), with normalized locally-time-averaged energy density and phase in the upper and lower panels respectively. The spot size at the focus is 4 μm (peak wavelength $\lambda_0 = 0.8$ μm) and the full-width-at-half-maximum (FWHM) pulse duration is 7.75 fs (2.9 cycles), corresponding to parameters $a = s = 123$. (d) shows the intensity of the pulse spectrum on the surface of the light cone, revealing a frequency spread (standard deviation) of $\Delta\omega/\omega_0 = 5.9\%$.

Throughout this paper (including Fig. 1c), we focus on the linearly-polarized TD wavepacket, which is given by

$$\begin{aligned}\mathbf{E} &= \nabla \times \nabla \times \mathbf{\Pi}_x \\ \mathbf{H} &= \varepsilon \frac{\partial}{\partial t} \nabla \times \mathbf{\Pi}_x\end{aligned}, \qquad (7)$$

where $\mathbf{\Pi}_x = \psi \hat{x}$. Notably, Eqs. (4) and (7) contain neither counter-propagating components nor singularities (proof in Supplementary Information (SI) Section 1, and derivation of Eq. (4) in SI Section 2). This makes Eqs. (4) and (7) the first finite-energy scalar and vector wavepackets (respectively) that are *closed-form* solutions of Maxwell's equations for an electromagnetic wavepacket that is free of approximations. We also find a fully analytical formula for the wavepacket in the spectral domain (SI Section 3) as plotted in Figs. 2d and 3f.

One reason that motivates the search for closed-form solutions of finite-energy electromagnetic pulses is the accurate modeling of single-cycle and sub-single-cycle pulses. The push towards shorter laser pulse durations have motivated scientists to seek non-paraxial descriptions for focused pulses in free space. This has been a challenge that attracted much research over more than 25 years [36-44]. Although few-cycle pulses are now regularly generated (and even experimentally demonstrated at high intensities) [45-48], all existing closed-form models for these pulses still suffer from drawbacks like the existence of points of divergence (where the field goes to infinity, which is non-physical) and backward propagating components. Eq. (4) and (7), on the other hand, describe forward-propagating, finite-energy pulses without any of these limitations. An example of a tightly-focused, few-cycle pulse modeled by (4) is given in Fig. 2. Due to the increasing popularity of intense, few-cycle pulses, this fully analytical description of such pulses can be helpful in understanding ultrafast and non-paraxial beam propagation phenomena that depart from the intuition of the many-cycle, paraxial regime.

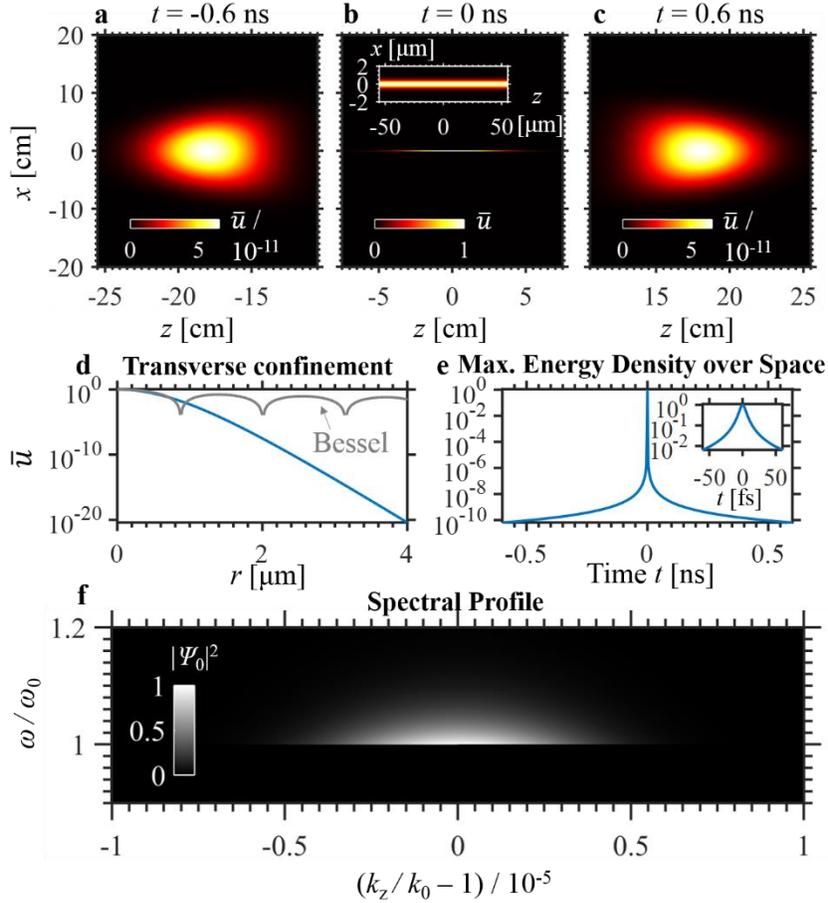

Figure 3. A linearly-polarized Abruptly Focusing Needle pulse. Snapshots of the propagating pulse are provided at $t = -0.6$ ns (a), $t = 0$ (b) and $t = 0.6$ ns (c). The spot size at the focus is 1.22 μm (peak wavelength $\lambda_0 = 0.8$ μm) and the needle length at the focal time is 6 cm, corresponding to a length:width aspect ratio of 50,000. For this case, $a = 12.3$, $s = 85 \times 10^9$. (d) shows transverse decay of the locally-time-averaged energy density at $t = 0$, revealing excellent transverse confinement that is superior to a Bessel beam of equivalent main lobe spot size. The abrupt focusing and defocusing nature of the AFN pulse is shown in (e). (f) depicts the intensity of the pulse spectrum on the surface of the light cone, revealing a frequency spread (standard deviation) of $\Delta\omega/\omega_0 = 3.8\%$, which gives a FWHM pulse duration of 9.7 fs

The parameter space of the TD wavepackets in Eq. (4) may be divided into two regimes: the standard regime, which describes a regular pulse, and the Abruptly Focusing Needle (AFN) regime, which describes a new kind of abrupt 4D spatiotemporal focusing of light into a strongly localized needle-shaped region. We define the AFN regime by the condition that the full-width-at-half-maximum (FWHM) length $L$ of the high intensity region at the focal time exceeds the nominal confocal parameter $k_0 w_0^2$, i.e., $L > k_0 w_0^2$, $w_0$ being the beam waist radius. An example of a wavepacket in the AFN regime is shown in Fig. 3, which shows an energy density that peaks strongly in a highly concentrated needle-shaped region (length:width aspect ratio of 50,000 in Fig. 3) near the axis ($r = 0$), at the focal time. The extreme aspect ratio can be attributed to the small spread in $k_z$ in Fig. 3f, in such a way that causes the different frequency components to constructively interfere only within the needle-shaped regime in a short window of time about the focal time. In fact, the aspect ratio can be made arbitrarily large by making the spread in $k_z$ arbitrarily small.

The plots in Fig. 3 highlight the abruptness of the temporal focusing and defocusing as well as the extreme spatial localization: the energy density decays by orders of magnitude as one moves just several cycles spatially or temporally from the focal region. Therefore, the AFN spatial localization has advantages over the Bessel-like or Airy-like transverse decay in conventional needle and non-diffracting beams. This is highlighted by the comparison with a Bessel beam's energy density profile in Fig. 3d. The transverse confinement and abruptness of focusing for other values of the parameter *a* are examined in Fig. 4. The AFN is especially useful in scenarios where one would like to avoid damaging the surrounding environment, because the spatiotemporal dynamics causes the pulse to abruptly focus on the target spot and abruptly defocus after it. When such dynamics is desirable, the rapid defocusing of the pulse gives it unique advantages over previous autofocusing beams that have secondary peaks after the focus [3]. For instance, the AFN can potentially be used to target tumors very precisely in cancer treatment, drill holes of very precise dimensions in laser machining, or trigger nonlinear processes (e.g., multiphoton absorption) in a well-defined region of 3D space.

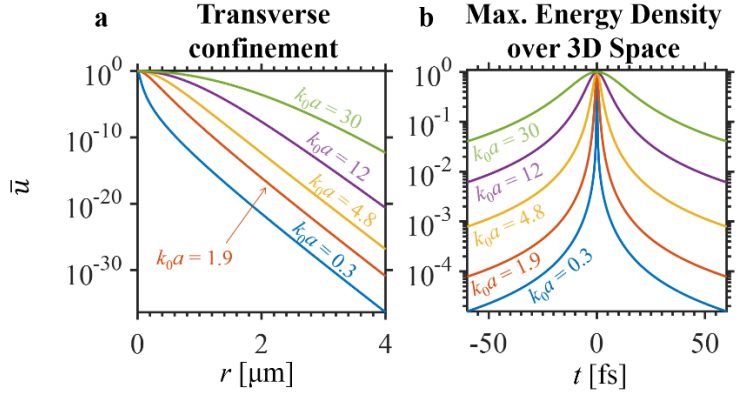

Figure 4. Strong localization in both space and time of linearly-polarized Abruptly Focusing Needle (AFN) pulses. The transverse energy density profiles at the focal plane for various *a* are shown in (a), where we see a rapid decay: exponential for $r \ll a$; for very large r's the decay is proportional to $r^{-2s-8}$, but by then the energy density has already fallen by tens of orders of magnitude. The maximum energy density over all space as a function of time is shown in (b). At small values of *a*, the peak energy density increases by many orders of magnitude as the pulse approaches the focus. As an example, the energy density of the $k_0 a = 0.3$ case increases by almost 5 orders of magnitude as the pulse traverses a mere distance of 18 μm in a time of 60 fs.

We give further examples of linearly-polarized and radially-polarized electromagnetic vector wavepackets in SI Section 1. Equation (4) can also be used to generate infinite new classes of solutions (e.g., counterparts of the paraxial Hermite-Gaussian, Laguerre-Gaussian etc. families) since, for instance, any linear combination of any multiplicity of partial derivatives in space and time of (4) is also a solution of the wave equation. Additional families are possible by substituting complex values for *s*. Importantly, any member of such a family constitutes a pulse of finite energy, unlike many conventional beams and pulses (e.g., plane waves, Airy beams [1], general accelerating beams [2], and Bessel beams [49]) that must be truncated in order to carry finite energy. Note that the most well-known existing analytical solution of an electromagnetic pulse, the conventional complex-source-point solution [36, 44], also carries finite energy. However, it consists of both forward and backward propagating components, which make it an approximate description (removing the backward propagating components results in a pulse with diverging amplitudes at singular points).

By applying the paraxial, many-cycle limit $k_0 a \gg 1$, $s \gg 1$, with $s \gg k_0 a$, we obtain the fundamental relation between the transverse width and pulse duration. Eq. (4) reduces to the result

$$\psi \approx \sigma \exp[ik_0(z-ct)]\exp\left[-\frac{k_0^2}{2s}(z-ct)^2\right], \tag{8}$$

with the beam envelope given by

$$\sigma \equiv \frac{1}{k_0(ct-ia)}\exp\left[\frac{ik_0 r^2}{2(ct-ia)}\right]. \tag{9}$$

Eq. (9) relates the FWHM pulse duration $\tau$ to the waist radius $w_0$ as

$$\tau = \frac{k_0 w_0^2}{c}. \tag{10}$$

A short pulse duration is thus associated with high transverse confinement, but any finite length:width aspect ratios in (8) are possible for any pulse duration and transverse width. Given that Eq. (2) is mathematically similar in form to the Helmholtz equation, it is not surprising that Eq. (10) is reminiscent of the relation between the confocal parameter and beam waist in the conventional Gaussian beam solution of the paraxial Helmholtz equation. For a beam waist diameter of 1.22 μm and wavelength 0.8 μm, Eq. (10) predicts a pulse duration of 9.7 fs, which agrees with the temporal FWHM obtained via exact numerical integration in Figs. 1 and 3.

Another noteworthy feature of Eq. (4) is that the velocity of the on-axis intensity-peak is always exactly luminal, a fact which holds regardless of pulse duration and focusing, and in spite of the superluminal on-axis phase velocity. This can be seen directly by setting $r = 0$ in (4), to give

$$\psi\big|_{r=0} = \frac{-i}{k_0(ict+a)}\left(\frac{1}{k_0(ict+a)}f_0^{-s-1} + \frac{s+1}{s}f_0^{-s-2}\right), \tag{11}$$

where $f_0 \equiv 1 - \mathrm{i} k_0 (z - \mathrm{c}t)/s$. From (11), we note that for any given $t$, the peak intensity always occurs at $z = \mathrm{c}t$, where the minimum of $|f_0|$ is located and where the parenthesized expression in (11) thus peaks. Hence, the on-axis peak intensity moves with speed c, even as the off-axis part of the pulse causes the overall pulse centroid to move slower than c. An illustration comparing these aspects of the TD wavepacket with those of the standard Gaussian beam solution is provided by Fig. S2.1, in Supplementary Information Section 2. Having the pulse peak maintaining the speed of light despite the inevitable diffraction is very useful for light-matter interactions taking place close to the laser beam axis, such as laser-driven particle acceleration [50, 51].

Pulse shaping in both space and time is widely used today to realize user-designed electromagnetic fields, although it can still be challenging especially for ultrashort pulses in the optical range [52, 33]. The use of pre-engineered phase masks and amplitude masks or of spatial light modulators (SLMs), in either real space or Fourier space, is probably the most viable option for realizing these pulses experimentally for optical frequencies [33]. New methods of spatiotemporal pulse shaping continue to be discovered and broaden the range of alternative approaches [53]. We next analyze a possible realization through time-dependent current distributions (Fig. 5) located far from the respective focal regions, that would generate the wavepackets studied in Figs. 2 and 3. Specifically, Fig. 5 shows the localized current distributions Re$\{\mathbf{J}_s\}$ in the x-y plane at a given $z$ that can produce the TD wavepackets of Figs. 2 and 3 in the far-field. The current distributions may be analytically determined via the expression $\mathbf{J}_s = \hat{z} \times \mathbf{H} = J_{sy}\hat{y}$ [35]. As we see in Fig. 5, current modulations on the order of the carrier period (2.7 fs here since we chose a central wavelength of 0.8 μm) are required. It may be possible to induce such current density modulations using an ordinary pulsed laser incident on a metasurface, where nano-antenna arrays are used to create the desired current distributions. Alternatively, since the Maxwell equations are scale-invariant, one can demonstrate a proof-of-principle electromagnetic TD microwave pulse, by driving an antenna array that supports the modulated electric current shown in Fig.5, after scaling the result to microwave frequencies.

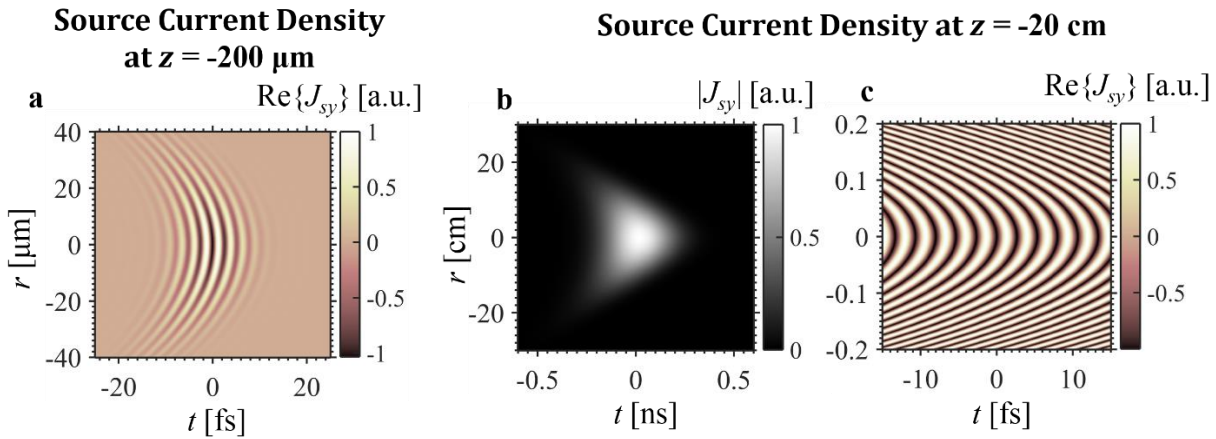

Figure 5. Surface current densities localized along a plane that are capable of generating the wavepackets shown in Fig. 2 (in (a)) and Fig. 3 (in (b) and (c)) respectively. (b) shows the magnitude of the current density instead of the current density itself as the variations are too fine to be resolved in this graphic. (c) shows the current density in a zoomed-in portion of (b).

We have introduced a new class of electromagnetic pulses that can create high intensity regions of arbitrary finite dimensions in free space, circumventing the tradeoff between length and narrowness typically enforced by Fourier optics. The pulses we introduce reveal the ability of Maxwell's equations to support abruptly focusing needles of light (More generally, the TD pulses we introduce are valid in linear, homogeneous, time-invariant and isotropic media and not just free space). We can directly infer this from our closed-form finite-energy pulse solutions of Maxwell equations. These pulses, presented in Eqs. (8) and (9), exhibit additional intriguing physical phenomena including an exactly luminal on-axis peak velocity. The ability to create highly localized intensity hotspots of any finite aspect ratio is promising for scenarios where one would like to avoid damaging surrounding environment, for instance, to minimize damage of healthy tissue surrounding cancer tissue. Potential applications range from high-precision pump-probe spectroscopy and laser-driven particle acceleration, to laser nanosurgery and high-aspect-ratio laser drilling of features like microfluidic channels.

**Supporting Information**: Please see online for Supplementary Information, where we further discuss the vector solutions of temporally diffracting electromagnetic wavepackets, present their derivation, obtain exact closed-form expressions for their corresponding spectra, and prove the absence of singularities in these solutions.

**Funding Information:** The work was supported by the Science and Engineering Research Council (SERC; grant no. 1426500054) of the Agency for Science, Technology and Research (A*STAR), Singapore. The research of I.K. was supported by the Seventh Framework Programme of the European Research Council (FP7–Marie Curie IOF) under grant agreement no. 328853 – MC–BSiCS.

**Acknowledgments:** The authors would like to acknowledge helpful discussions with Prof. Mordechai Segev and Dr. Yaakov Lumer.